# High density nonmagnetic cobalt in thin films

Nasrin Banu[1], Surendra Singh[2], Saibal Basu[2], Anupam Roy[3], Hema C. P. Movva[3], B. N. Dev[1]

[1] Department of Materials Science, Indian Association for the Cultivation of Science, 2A & 2B Raja S. C. Mullick Road, Jadavpur, Kolkata-700032, India

[2] Solid State Physics Department, Bhabha Atomic Research Centre, Mumbai 400085, India.

[3]Microelectronic Research Center, The University of Texas at Austin,10100 Burnet Road, Bldg 160, MER 1.606J, Austin, Texas 78758, USA.

**Abstract**

Recently high density (HD) nonmagnetic (NM) cobalt has been discovered in a cobalt thin film, grown on Si(111). This cobalt film had a natural cobalt oxide at the top. The oxide layer forms when the film is taken out of the electron-beam deposition chamber and exposed to air. Thin HD NM cobalt layers were found near the cobalt/silicon and the cobalt-oxide/cobalt interfaces, while the thicker mid-depth region of the film was hcp cobalt with normal density and normal magnetic moment. If an ultrathin film of gold is grown on the cobalt layer, before exposing it to air, the oxidation of the cobalt surface layer is prevented. It is important to investigate whether the growth of HD NM cobalt layers in the thin film depends on (i) capping of the film by the gold layer, (ii) the film thickness and (iii) the nature of the substrate. The results of such investigations, presented here, indicates that for cobalt films capped with a thin gold layer, and for various film thicknesses, HD NM cobalt layers are still observed. However, instead of a Si substrate, when the cobalt films are grown on oxide substrates, such as silicon oxide or cobalt oxide, HD NM cobalt layers are not observed.

## 1. Introduction:

Magnetism in dense phases of 3d transition metal elements or compounds is of interest because of their existence in the Earth's core, where materials exist under high pressure. In the laboratory, such dense phases are obtained by applying pressure to bulk materials. However, materials in the form of thin films may show unusual structures and properties which can be obtained in the bulk form only under extreme conditions, such as, high pressure and/or high temperature [1]. For example, for bulk cobalt, fcc structure exists at temperatures above 417 $^oC$ [2] or under high pressure [3]. On the other hand, in the form of thin films cobalt may exist in fcc structure at room temperature and normal pressure and their properties vary widely compared to bulk cobalt [1]. Bulk Bi is superconducting under high pressure [$T_c$ about 8 K] [4]. On the other hand thin films of Bi is superconducting at ambient pressure ($T_c$ about 5-6 K) [5]. (Very recently superconductivity in bulk Bi was discovered, with a $T_c$ of about 0.5 millikelvin at ambient pressure [6]). Thin films provide extra degrees of freedom which may change the thermodynamic variables required to get phase transitions in thin films. Recently we have discovered a high density nonmagnetic fcc phase of cobalt in a cobalt thin film [7-9]. For bulk cobalt this is expected to happen only under high pressure.

Extensive investigations using various experimental techniques confirmed the existence of high density (HD) nonmagnetic (NM) cobalt layers, a few nanometer (nm) thick, near the top and the bottom of a 25 nm Co thin film on Si [8, 9]. The Co film in this case, grown of Si, had a natural oxide on the surface. The film configuration was CoO/Co/Si. HD NM cobalt layers were also observed when the Co layer was protected by a thin Au capping layer [7]. The Au capping layer prevents the formation of CoO on the cobalt film. In order to investigate if the film thickness has any influence on the formation of HD NM Co, we have investigated several films of different thicknesses and with a Au capping layer. The behavior of formation of HD NM layers remains practically unaltered for the investigated range of film thickness. We also addressed another question. Why was the HD NM Co not found in earlier investigations in Co thin films? Radu et al. have carried out experiments on Co films grown on $Al_2O_3$ substrates [10]. In the work of Velthuis et al. [11], Co films were claimed to have been grown on Si. However, it was not mentioned whether the native oxide layer on Si was removed prior to Co deposition. Presumably, the Co layer was grown on silicon oxide. As the Co thin films in earlier investigations were grown on oxides [10, 11], in order to investigate the substrate dependence of HD NM Co growth we carried out experiments on Co films grown on oxides ($SiO_2$, CoO) as well. The HD NM phase of Co has not formed in these cases. Here we present the results of these investigations.

## 2. Experiments

We have deposited all the films by electron beam deposition at room temperature under high vacuum condition. The samples are capped with gold to prevent oxidation of the film surface. The capping gold layer is also deposited by e-beam evaporation following the deposition of the cobalt film without breaking vacuum. We have prepared the following samples: Au(2nm)/Co(25nm)/Si(111), Au(2nm)/Co(22nm)/Si(111), Au(2nm)/Co(12nm)/Si(111), Au(2nm)/Co(4nm)/Si(111) and Au(2nm)/Co(25nm)/$SiO_2$/Si(111). Film thicknesses mentioned here are nomonal thicknesses. Si(111) substrates were cleaned by piranha solution followed by HF etching. Co films were grown on these Si(111) substrates. In the last case a $SiO_2$ layer was grown on Si(111) prior to Co deposition. After the deposition of the film and the capping layer the samples were taken out of the vacuum chamber for different measurements. We have prepared one sample where double layers of [CoO/Co] were grown on Si(111). After deposition of one Co layer, it was exposed to air to form a CoO layer on it. Then this sample was inserted into the deposition chamber for deposition of another Co layer. The sample was then taken out and exposed to air for the formation of the second CoO layer. This sample structure is CoO/Co/CoO/Co/Si(111). We have carried out X-ray reflectometry (XRR) [12] and polarised neutron reflectometry (PNR) [12,13] for characterization. XRR was carried out with an 18 kW Cu Kα source and PNR was carried out with polarized neutron of wavelength 2.5 Å, obtained from the DHRUVA reactor at BARC, Mumbai. Secondary ion mass spectrometry (SIMS) measuremnets were carried out at the Surface Physics Division, SINP, Kolkata with 4 keV Ar$^+$ ions and a beam current of 20 nA.

## 3. Results and discussion

XRR and PNR techniques provide depth dependent structure of the sample with sub-nanometer depth resolution averaged over the lateral dimensions of the entire sample (typically 100 mm$^2$) [12,13]. XRR and PNR involve measurement of the x-ray/neutron radiation reflected from a sample as a function of wave vector transfer $Q$ (i.e., the difference between the outgoing and incoming wave vectors). In case of specular reflectivity $Q = \frac{4\pi}{\lambda} sin\theta$, where θ is the angle of incidence and λ is the wavelength of x-ray/neutron, and it is qualitatively related to the square of the Fourier transform of the scattering length density (SLD) depth profile $\rho(z)$ (normal to the film surface) [12,13]. For XRR, $\rho_x(z)$ is proportional to electron density whereas for PNR, $\rho(z)$ consists of nuclear and magnetic SLDs such that

$$\rho^{\pm}(z) = \rho_n(z) \pm CM(z) \qquad (1)$$

where $C$ = 2.9109×10$^{-9}$ Å$^{-2}$ m/kA, and $M(z)$ is the magnetization (kA/m) depth profile [12,13]. The sign +(-) is determined by the condition when the neutron beam polarization is parallel (antiparallel) to the applied field. The corresponding reflectivities are denoted as $R^{\pm}$. The results of XRR and PNR measurements and the analyses for different thin film structures are presented below.

### 3.1 CoO/Co(25 nm)/Si(111)

For the sake of comparison, here we reproduce the results for a Co film without capping. If there is no capping on the Co film and the film is exposed to air, a thin cobalt oxide layer is formed on the top of the Co film. So the film structure is CoO/Co/Si. The XRR data from this sample are shown in Fig. 1. The electron scattering length density (ESLD) depth profile, shown in the inset, provides the best fit to the XRR data. ESLD is proportional to the actual material density. The ESLD or the density depth profile indicates the presence HD cobalt near the CoO/Co and the Co/Si interfaces. The possibility of this HD material being anything else other than Co has been ruled out via other experiments [9].

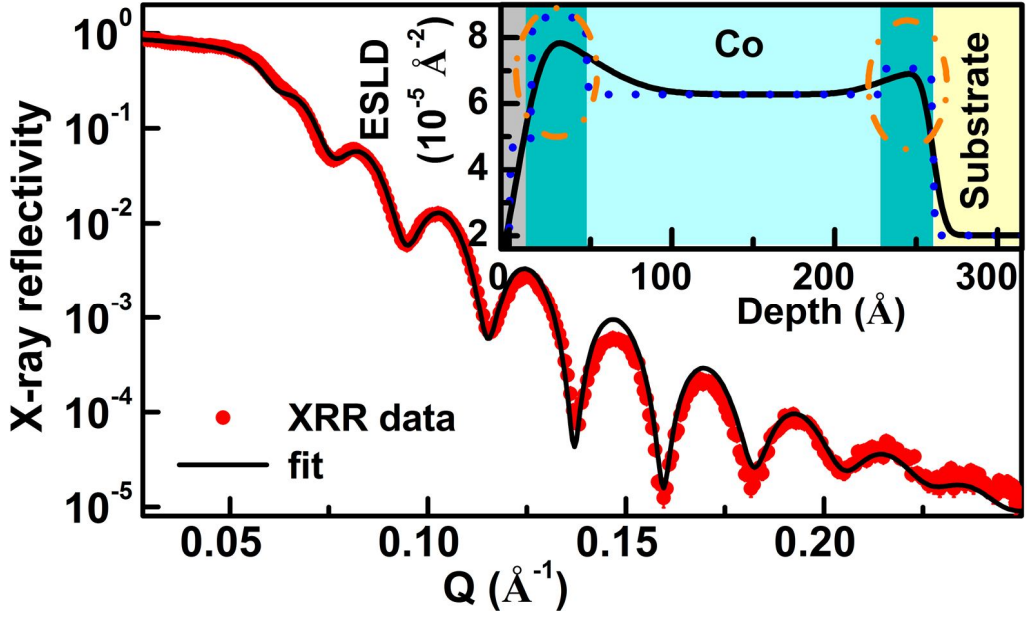

Fig. 1 XRR data (circle) and the fitted curve (solid line). The ESLD depth profile (solid line), shown in the inset, provides the best fit. HD cobalt regions (deep green) are marked by ovals in the inset. The ESLD depth profile shown by the dotted line (histogram) does not include the effect of interface roughness. (From Ref. 9).

Table-I: Parameters, such as layers, their thicknesses, ESLDs and surface/interface roughness obtained from the analysis of the XRR data.

| Layer | Thickness (Å) | ESLD ($10^{-5}$ Å$^{-2}$) | Roughness (Å) |
|---|---|---|---|
| CoO | 13±2 | 4.68±0.10 | 7±2 |
| HD Co | 35±3 | 8.56±0.20 | 8±2 |
| Co | 183±5 | 6.28±0.10 | 24±4 |
| HD Co | 34±3 | 7.20±0.15 | 13±3 |
| Si substrate | - | 2.05±0.05 | 5±2 |

PNR results also provide the density depth profile like XRR. However, PNR additionally provides the magnetic depth profile [12,13]. PNR results are shown in Fig. 2. Instead of ESLD, in PNR one obtains nuclear SLD (NSLD), which is also proportional to the mass density. PNR additionally provides magnetic SLD (MSLD) depth profile. The nuclear SLD and the magnetic SLD depth profile in Fig. 2(b) provide the best fit (Fig. 2(a)) to the PNR data. PNR results also confirms the existence of HD layers of Co. We notice that MSLD (or the magnetic moment density) goes nearly to zero in the HD Co regions. HD NM Co layers are highlighted in Fig. 2(b) by a deep green shade. The model of a cobalt layer with normal density and normal magnetic moment density over the whole depth of the film (Fig.2(d)) does not fit the PNR data, as seen in Fig. 2(c).

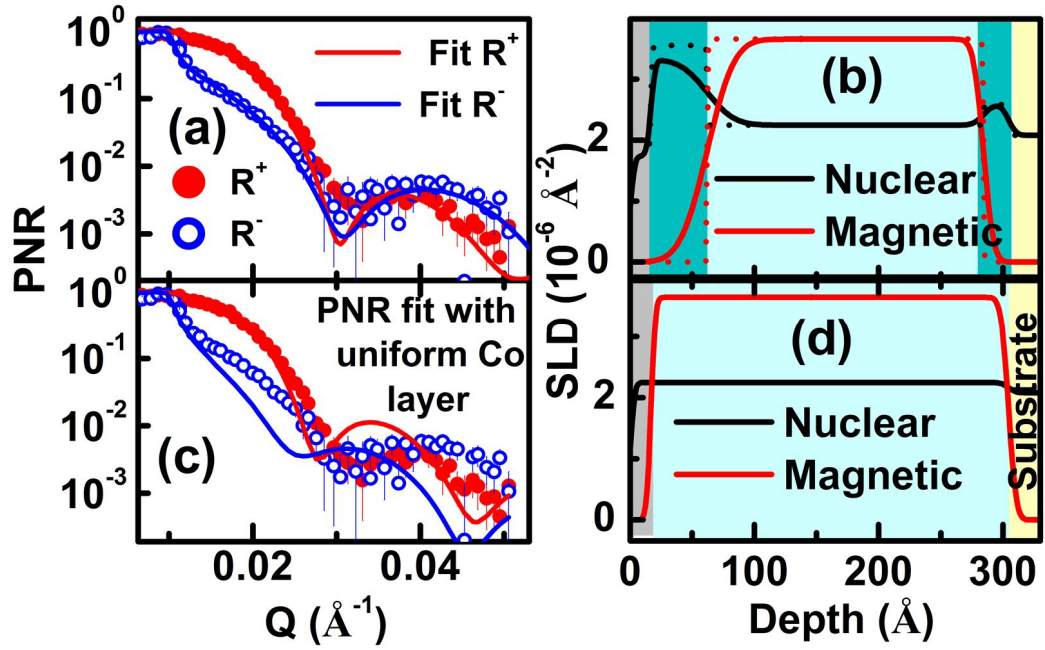

Fig. 2 (a) PNR data (circles) and the fitted curves (solid lines). (b) The nuclear and the magnetic SLD depth profiles (solid lines) produce the best fit to the data in (a). The dotted lines (histograms) in (b) represent the depth profiles without including the interface roughness effects. (c) PNR data and the fitted curves corresponding to the ESLD and MSLD depth profiles in (d). (From Ref. 9).

Table-II: Different layers and their thicknesses, NSLD, MSLD and roughness parameters, as obtained from the analysis of the PNR data.

| Layer | Thickness (Å) | NSLD ($10^{-6}$ Å$^{-2}$) | MSLD ($10^{-6}$ Å$^{-2}$) | Roughness (Å) |
|---|---|---|---|---|
| CoO | 17±2 | 4.03±1.00 | 0 | 14±2 |
| HD Co | 45±3 | 3.57±0.20 | 0 | 3±2 |
| Co | 222±6 | 2.30±0.10 | 3.65±0.12 | 17±3 |
| HD Co | 20±2 | 2.64±0.12 | 0 | 8±2 |
| Si sub | - | 2.09±0.07 | 0 | 5±1 |

The deviation between the values of the parameters (compare Tables I and II), extracted from XRR and PNR data arises mainly for two reasons: (i) XRR results represent the parameter values averaged over an area of 100 mm$^2$ (sample size 1x1 cm$^2$) whereas the PNR results represent an average over 2500 mm$^2$ (sample size 5x5 cm$^2$), and (ii) data obtained over a large Q-range in XRR provides better precision compared to the PNR data (Q-range smaller by a factor of about 5).

## 3.2 Cobalt films with Au capping and the dependence of HD NM cobalt formation on film thickness

### (i) Au(2 nm)/Co(25 nm)/Si(111)

Before the sample was taken out of the vacuum deposition chamber, a thin Au layer was deposited on the Co film, so that surface oxidation of the Co film was prevented when it was exposed to air. XRR and PNR data and the ESLD, NSLD and MSLD depth profiles obtained from this sample are shown in Fig. 3. Table III(a) lists various parameters for different layers obtained from the best fit of the XRR data. Table III(b) shows the corresponding parameters obtained from the analysis of the PNR data. We still observe the same features of HD NM cobalt at the top of the cobalt film, just below the Au layer, and at the Co/Si interface.

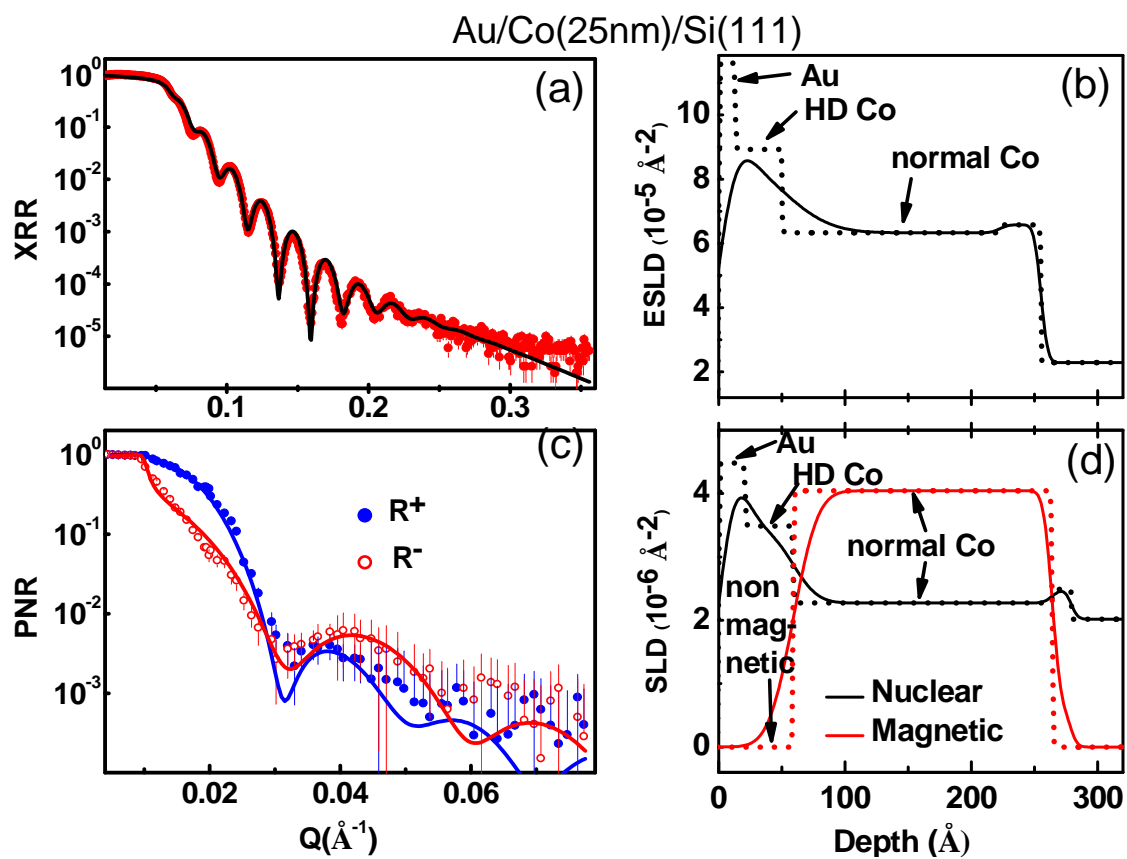

Fig. 3 (a) XRR data (circle) and the fitted curve (solid line). (b) The ESLD depth profile (solid line) that produces the best fit in (a). The dotted line (histogram) in (b) represents the depth profile without the interface roughness. (c) PNR data (circle) and the fitted curves (solid line). (d) ESLD and MSLD depth profiles (solid lines) that provide the best fits shown in (c). The dotted lines (histogram) in (d) represent the depth profiles without including the interface roughness.

Table-III(a): Parameters, such as layers, their thicknesses, ESLDs and surface/interface roughness obtained from the analysis of the XRR data.

| Layer | Thickness (Å) | ESLD ($10^{-5}$ Å$^{-2}$) | Roughness (Å) |
|---|---|---|---|
| Au | 13±2 | 11.7±0.30 | 10±2 |
| HD Co | 35±3 | 8.97±0.20 | 12±3 |
| Co | 170±5 | 6.40±0.30 | 20±4 |
| HD Co | 32±3 | 6.65±0.35 | 5±2 |
| Si sub | - | 2.08±0.05 | 5±2 |

Table-III(b): Different layers and their thickness, NSLD, MSLD and surface/interface roughness parameters, as obtained from the analysis of the PNR data.

| Layer | Thickness (Å) | NSLD ($10^{-6}$ Å$^{-2}$) | MSLD ($10^{-6}$ Å$^{-2}$) | Roughness (Å) |
|---|---|---|---|---|
| Au | 17±3 | 4.50±0.25 | 0 | 10±2 |
| HD Co | 35±3 | 3.50±0.20 | 0 | 11±3 |
| Co | 195±10 | 2.26±0.12 | 4.04±0.27 | 17±5 |
| HD Co | 20±5 | 2.50±0.25 | 0 | 5±2 |
| Si sub | - | 2.05±0.05 | 0 | 5±2 |

For the Au-capped Co films we have also investigated a few other films of different thicknesses, namely 22 nm, 12 nm and 4 nm. These cases are presented below and the results are shown in Fig. 4, 5 and 6 respectively.

## (ii) Au/Co(22 nm)/Si(111)

XRR and PNR results for a 22 nm Co film are shown in Fig. 4 and Tables IV(a) and IV(b).

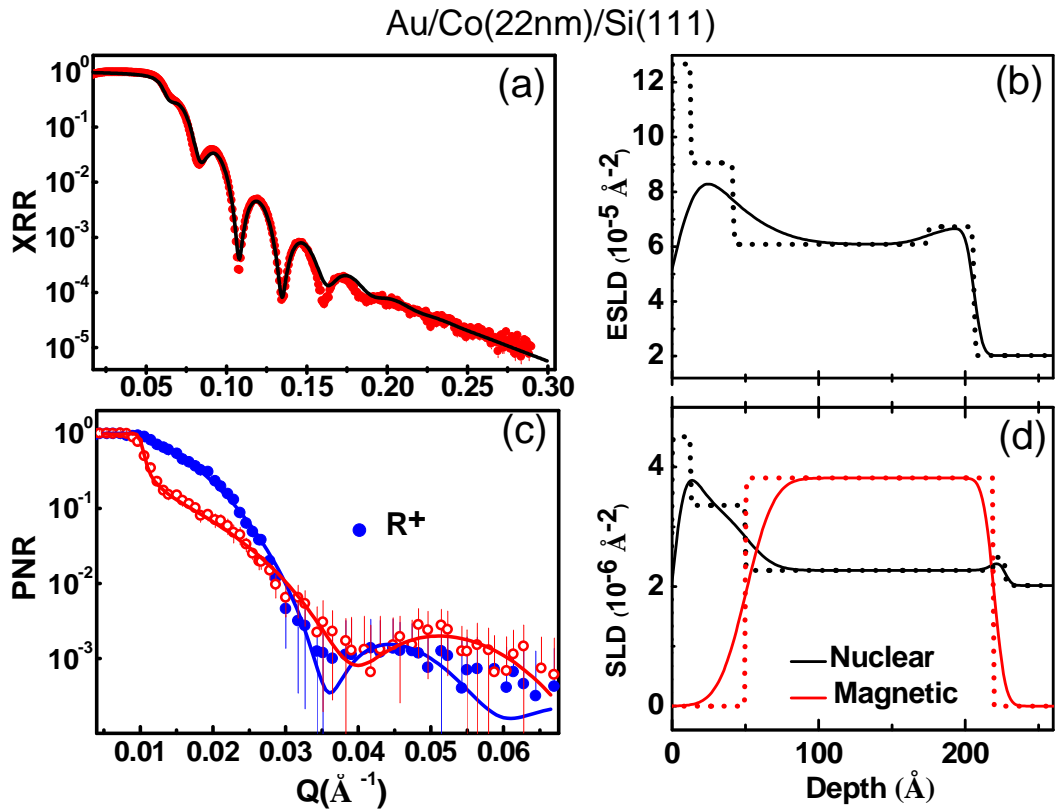

Fig. 4 XRR data (circle) and the fitted curve (solid line). (b) The ESLD depth profile (solid line) that produces the best fit in (a). (c) PNR data (circle) and the fitted curves (solid line). (d) ESLD and MSLD depth profiles (solid line) that provide the best fits shown in (c). The dotted lines (histogram) show the depth profiles without taking into account the effect of surface and interface roughness.

Table IV(a): Parameters, such as layers, their thicknesses, ESLDs and surface/interface roughness obtained from the analysis of the XRR data.

| Layer | Thickness (Å) | ESLD ($10^{-5}$ Å$^{-2}$) | Roughness (Å) |
|---|---|---|---|
| Au | 13±2 | 11.99±0.30 | 11±2 |
| HD Co | 30±3 | 8.75±0.20 | 17±3 |
| Co | 130±5 | 6.20±0.30 | 20±4 |
| HD Co | 32±3 | 6.75±0.35 | 11±2 |
| Si sub | - | 2.06±0.06 | 5±2 |

Table IV(b): Different layers and their thicknesses, NSLD, MSLD and surface/interface roughness parameters, as obtained from the analysis of the PNR data.

| Layer | Thickness (Å) | NSLD ($10^{-6}$ Å$^{-2}$) | MSLD ($10^{-6}$ Å$^{-2}$) | Roughness (Å) |
|---|---|---|---|---|
| Au | 13±3 | 4.50±0.25 | 0 | 10±2 |
| HD Co | 35±3 | 3.60±0.20 | 0 | 11±3 |
| Co | 160±10 | 2.25±0.12 | 3.85±0.25 | 17±5 |
| HD Co | 15±5 | 2.50±0.25 | 0 | 5±2 |
| Si sub | - | 2.05±0.05 | 0 | 5±2 |

### (iii) Au/Co(12 nm)/Si(111)

XRR and PNR results for a 12 nm Co film are shown in Fig. 5 and Tables V(a) and V(b).

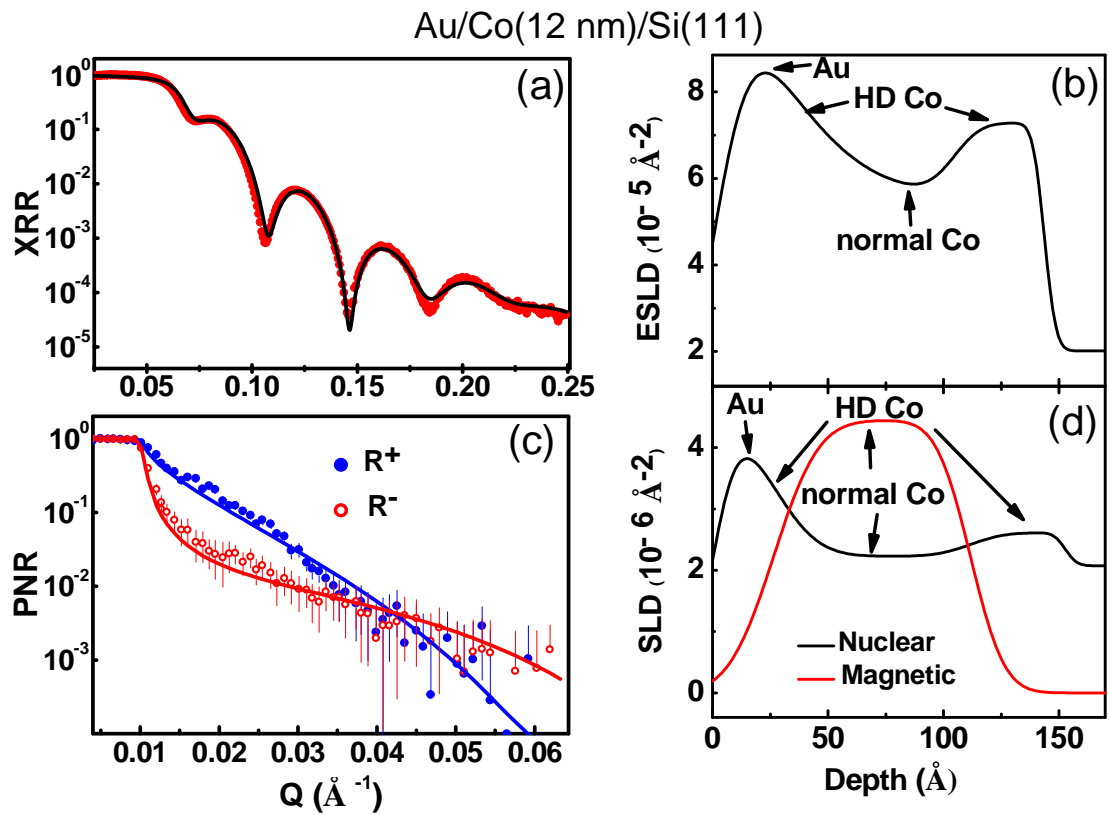

Fig. 5 XRR data (circle) and the fitted curve (solid line). (b) The ESLD depth profile that produces the best fit in (a). (c) PNR data (circle) and the fitted curves (solid line). (d) ESLD and MSLD depth profiles that provide the best fits shown in (c).

Table V(a): Parameters, such as layers, their thicknesses, ESLDs and surface/interface roughness obtained from the analysis of the XRR data.

| Layer | Thickness (Å) | ESLD ($10^{-5}$ Å$^{-2}$) | Roughness (Å) |
|---|---|---|---|
| Au | 20±2 | 11.7±0.30 | 12±2 |
| HD Co | 25±3 | 8.57±0.20 | 12±3 |
| Co | 60±5 | 6.00±0.30 | 22±4 |
| HD Co | 36±3 | 7.35±0.25 | 8±2 |
| Si sub | - | 2.07±0.07 | 5±2 |

Table V(b): Different layers and their thicknesses, NSLD, MSLD and surface/interface roughness parameters, as obtained from the analysis of the PNR data.

| Layer | Thickness (Å) | NSLD ($10^{-6}$ Å$^{-2}$) | MSLD ($10^{-6}$ Å$^{-2}$) | Roughness (Å) |
|---|---|---|---|---|
| Au | 21±3 | 4.50±0.25 | 0 | 9±2 |
| HD Co | 25±3 | 3.70±0.20 | 0 | 11±3 |
| Co | 62±4 | 2.24±0.12 | 4.42±0.17 | 17±5 |
| HD Co | 40±3 | 2.70±0.25 | 0 | 8±2 |
| Si sub | - | 2.05±0.05 | 0 | 5±2 |

**(iv) Au/Co(4 nm)/Si(111)**

XRR and PNR results for a 4 nm Co film are shown in Fig. 6 and Tables VI(a) and VI(b).

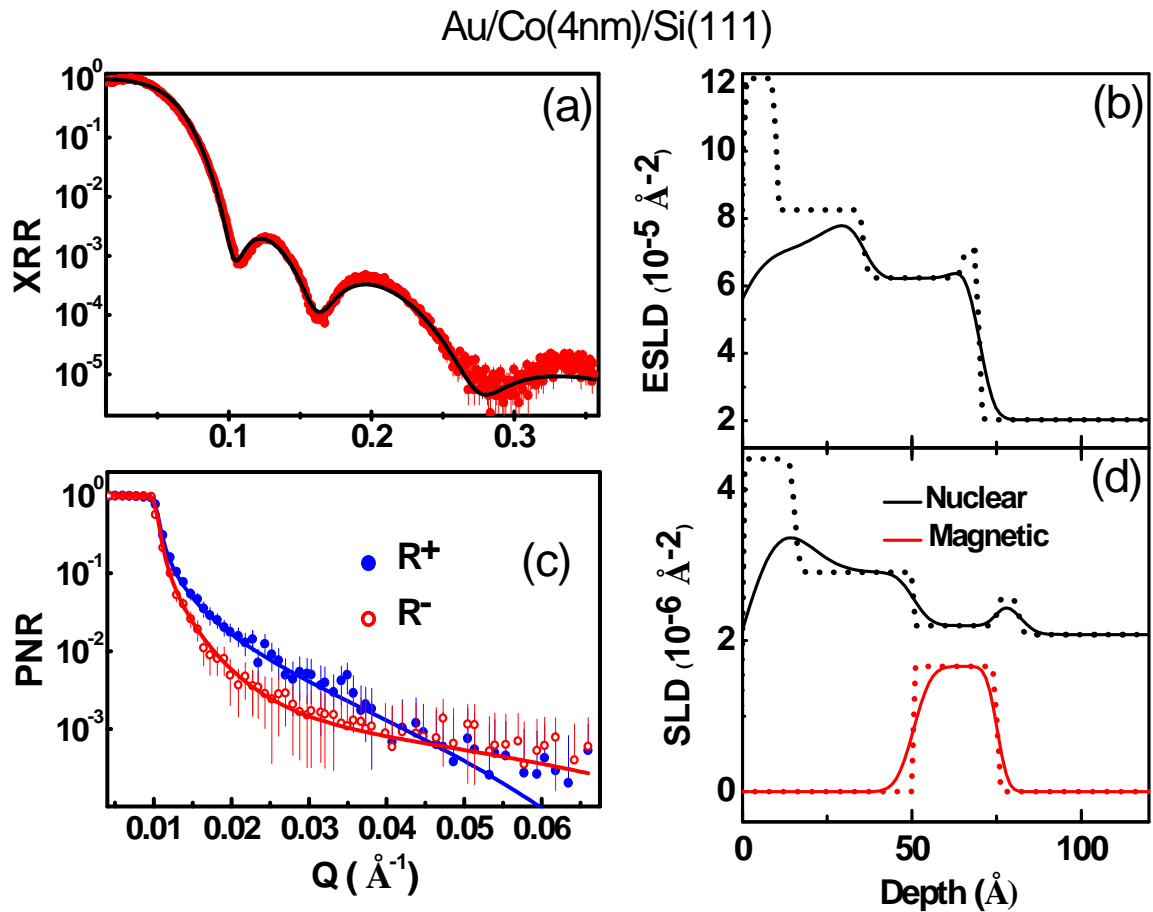

Fig. 6 XRR data (circle) and the fitted curve (solid line). (b) The ESLD depth profile that produces the best fit in (a). The dotted line (histogram) shows the depth profile without taking the roughness into account. (c) PNR data (circle) and the fitted curves (solid line). (d) ESLD and MSLD depth profiles that provide the best fits shown in (c). The dotted line profiles (histogram) do not include surface/interface roughness.

Table VI(a): Parameters, such as layers, their thicknesses, ESLDs and surface/interface roughness obtained from the analysis of the XRR data.

| Layer  | Thickness (Å) | ESLD ($10^{-5}$ Å$^{-2}$) | Roughness (Å) |
|--------|---------------|---------------------------|---------------|
| Au     | 10±2          | 11.7±0.30                 | 10±2          |
| HD Co  | 25±3          | 8.17±0.20                 | 8±2           |
| Co     | 25±4          | 6.20±0.20                 | 4±1           |
| HD Co  | 8±3           | 6.62±0.25                 | 5±2           |
| Si sub | -             | 2.06±0.05                 | 5±2           |

Table VI(b): Different layers and their thicknesses, NSLD, MSLD and surface/interface roughness parameters, as obtained from the analysis of the PNR data.

| Layer | Thickness (Å) | NSLD ($10^{-6}$ Å$^{-2}$) | MSLD ($10^{-6}$ Å$^{-2}$) | Roughness (Å) |
|---|---|---|---|---|
| Au | 12±2 | 4.40±0.20 | 0 | 10±2 |
| HD Co | 30±3 | 3.20±0.15 | 0 | 9±2 |
| Co | 25±4 | 2.22±0.12 | 1.74±0.17 | 4±1 |
| HD Co | 9±3 | 2.50±0.25 | 0 | 4±1 |
| Si sub | - | 2.06±0.05 | 0 | 5±2 |

Ideally, the ESLD and the NSLD depth profiles should be identical for a given sample. However, in reality for a thin film of large area, some lateral inhomogeneity is expected. As the XRR and the PNR measurements have provided information averaged over a sample area of about 1 cm$^2$ and 25 cm$^2$ respectively, some variation in the ESLD and NSLD depth profiles is expected. This is seen in all cases. The deviation is the most pronounced for the case of the thinnest film (4 nm), as seen from Fig. 6. In the depth profile of Fig. 6(b), the higher density of Au compared to HD Co, although clear from the histogram, is not seen in the profile that includes the effect of roughness. This is because the surface roughness of the Au film (Table VI(a)) is large and equal to its thickness; in addition, the roughness at the Au/HD-Co interface (8 Å, Table VI(a)) is also very large.

Compared to the other cases, we notice that for the 4 nm Co film (section 3.2(iv)) the MSLD value for the mid-depth region of the Co film, i.e. for normal Co, is much smaller. It is indeed much smaller than the value expected for saturation magnetization of normal Co. Prior to the PNR measurements, a couple of samples (25 nm and 22 nm films) were investigated by superconducting quantum interference device (SQUID) magnetometry. We found an in-plane magnetic anisotropy in these samples. We identified the easy axis of magnetization. During PNR measurement the applied magnetic field (1.7 kOe) on the sample was along this direction. Having completed PNR measurements on all samples, we investigated the thinnest sample (4 nm) by SQUID and found that the in-plane easy axis of magnetization is different from the assumed easy axis based on SQUID measurement on the thicker samples. That means that during the PNR experiment, the magnetic field was not applied along the easy axis for the 4 nm Co film and consequently the applied field could not saturate the magnetization. This is apparently the reason for the observed low MSLD value (Fig. 6(d) and Table VI(b)).

Results in section 3.2 show that the capping of the Co film by Au has no significant effect on the formation of the high density nonmagnetic Co layers within the Co thin film. As the surface oxidation (section 3.1) or Au-capping (section 3.2) occurred after the growth of the Co film had been completed, it is clear that the HD NM phase of Co has formed during the film growth and neither surface oxidation nor Au-capping has an effect on this growth. Also, irrespective of the film thickness, HD NM Co layers are always formed near the bottom and the top of the Co layer.

### 3.3. Substrate dependence

### (i) Co films grown on $SiO_2$ : Au(2 nm)/Co(25 nm)/$SiO_2$/Si(111)

In this case, a $SiO_2$ layer was first grown on Si(111) prior to Co deposition. XRR and PNR results are shown in Fig. 7 and Tables VII(a) and VII(b).

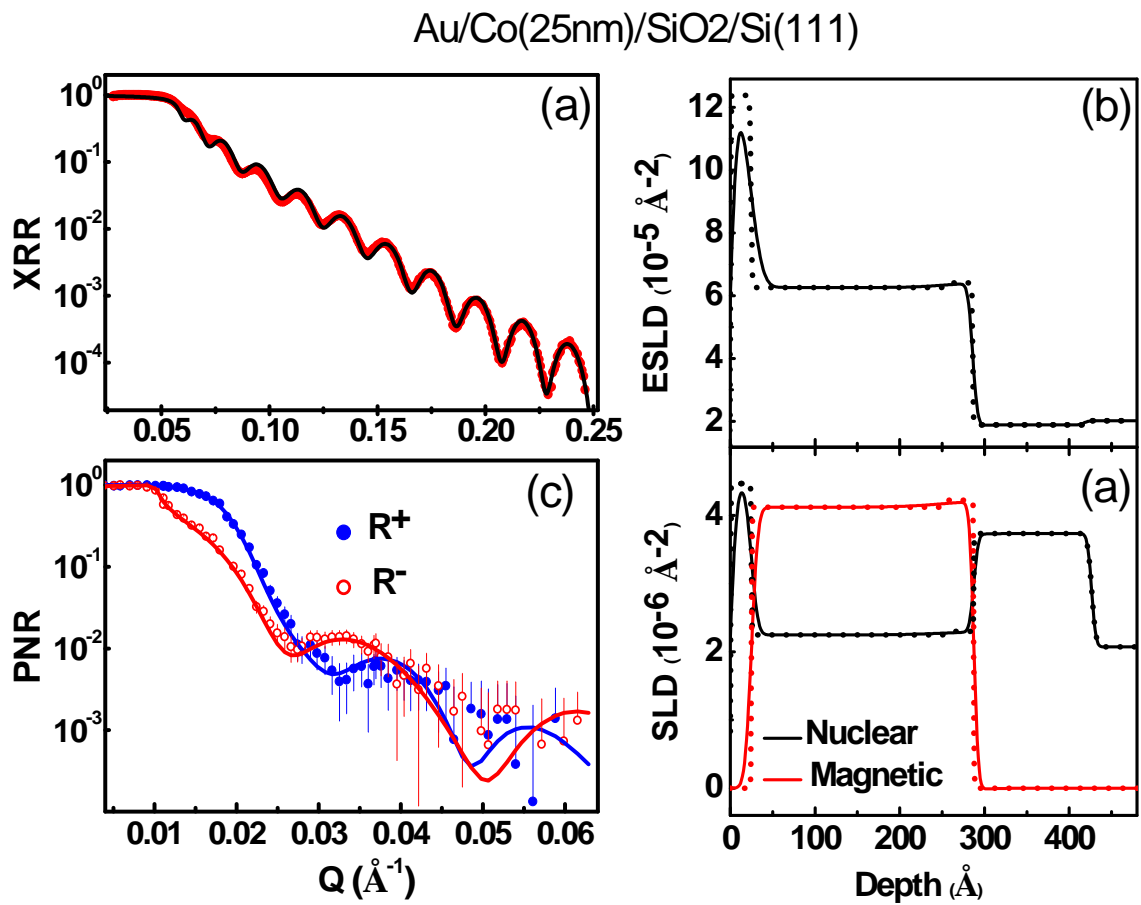

Fig. 7 (a) XRR data (circle, red) and the fitted curve (line, black). (b) The ESLD depth profile (solid line) that produces the best fit in (a). (c) PNR data (circle) and the fitted curves (solid line). The dotted-line depth profile (histogram) does not include the effect of roughness. (d) NSLD and MSLD depth profiles that provide the best fits shown in (c). The dotted-line depth profiles (histogram) do not include the effect of roughness.

Comparing Fig. 7(b) and Fig. 7(d), we notice that the major difference between the ESLD and the NSLD depth profiles appears in the $SiO_2$ film region. This is because the ESLD values for $SiO_2$ and Si (below the $SiO_2$ film) are comparable, while the NSLD value for $SiO_2$ is much larger than that of Si (see Tables VII(a) and VII(b)). The Co layer shows the usual MSLD, as expected from hcp Co. MSLD from Au, $SiO_2$ and Si layers is zero as they do not have magnetic moment. The results show that no high density nonmagnetic Co layers have formed within the Co films when the film is grown on $SiO_2$.

Table VII(a): Parameters, such as layers, their thicknesses, ESLDs and surface/interface roughness obtained from the analysis of the XRR data.

| Layer | Thickness (Å) | ESLD ($10^{-5}$ Å$^{-2}$) | Roughness (Å) |
|---|---|---|---|
| Au | 23±2 | 12.00±0.40 | 7±2 |
| Co | 260±12 | 6.35±0.30 | 12±3 |
| SiO$_2$ | 130±8 | 1.92±0.10 | 5±2 |
| Si sub | - | 2.10±0.10 | 5±2 |

Table VII(b): Different layers and their thickness, NSLD, MSLD and surface/interface roughness parameters, as obtained from the analysis of the PNR data.

| Layer | Thickness (Å) | NSLD ($10^{-6}$ Å$^{-2}$) | MSLD ($10^{-6}$ Å$^{-2}$) | Roughness (Å) |
|---|---|---|---|---|
| Au | 23±3 | 4.48±0.25 | 0 | 7±2 |
| Co | 263±10 | 2.24±0.12 | 4.10±0.27 | 11±4 |
| SiO$_2$ | 136±7 | 3.70±0.25 | 0 | 5±2 |
| Si sub | - | 2.04±0.05 | 0 | 5±2 |

**(ii) Comparison between film growth on SiO$_2$ and Si**

Yield of Co ions from SIMS measurements from two samples are shown in Fig. 8. For the film grown on the oxide substrate, the middle flat part is from the uniform Co film. The increased yield near the Au/Co and the Co/SiO$_2$ interfaces are due to matrix effect. For the film grown on Si, just below the Au/Co interface, the Co yield is higher compared to the case of Co film on SiO$_2$. This is highlighted in the inset. Near the Co/Si interface a small increase of Co yield is expected for two reasons, namely, for HD Co and the presence of a very small amount of CoSi (see SIMS in the supplementary material, Ref. 9) may have given the peak because of matrix effect. For the Co film on SiO$_2$ a peak appears at the Co/SiO$_2$ interface due to the matrix effect.

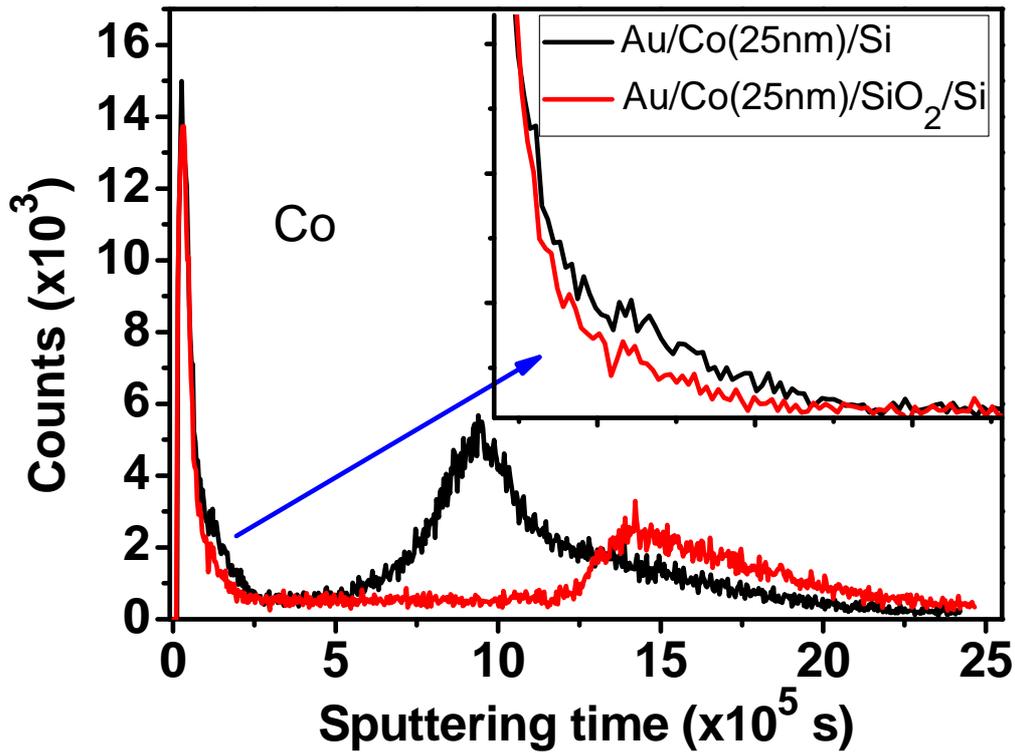

Fig. 8 Co yield in SIMS experiment on Co films grown on Si and on SiO$_2$.

**(iii) [CoO/Co]$_2$/Si(111)**

In this example of [CoO/Co/CoO/Co] layers, the first Co layer is grown on Si(111) and the second Co layer is grown on CoO. That means, within the same sample we investigate the substrate dependence of the HD Co growth. XRR and PNR results are shown in Fig. 9(a,c). Fig. 9(b) shows the ESLD depth profile. It is clear that HD Co layers are formed in the film grown on Si(111). However, HD Co is not formed in the Co film grown on CoO. So, Co films grown on the oxides, either SiO$_2$ (discussed in 3.3(i)) or CoO, do not show any HD Co formation. This is consistent with earlier investigations where Co films were grown on various oxide substrates [10,11].

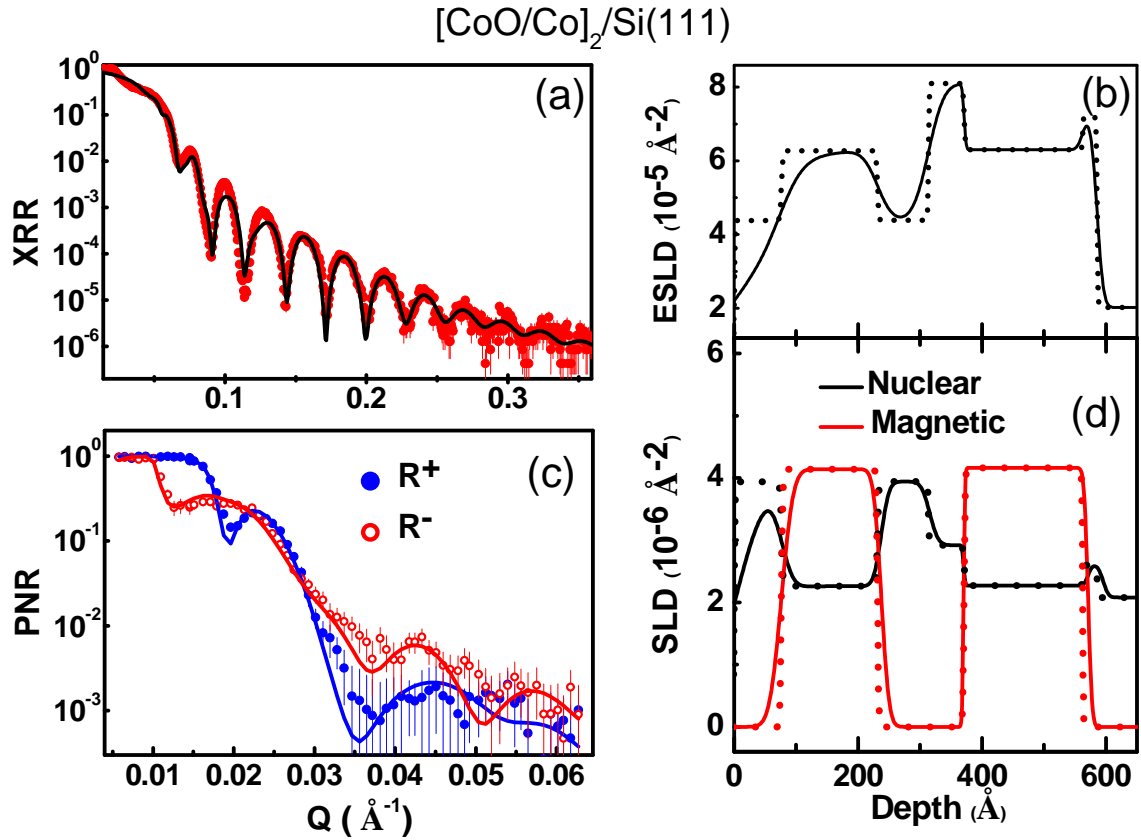

Fig. 9 (a) XRR data and the fitted curve. (b) The ESLD depth profile that produces the best fit. The dotted histogram represents the profile without taking into account the surface and interface roughness. (c) PNR data (filled and open circles) and the fitted curves (solid lines) (d) NSLD and MSLD depth profiles that provide the best fits. The dotted histograms show the depth profile without taking into account surface and interface roughness.

In the ESLD depth profile, the values of ESLD for the outer CoO and Co layers appear to be smaller than those for the inner CoO and Co layers. This is mainly the effect of much larger roughness of the outer CoO (72 Å) and Co (27 Å) layers (see Table IX(a)). PNR results are shown in Fig. 9(c) and Table IX(b). As in the XRR results, PNR results also show that the outer CoO and Co lyers have much larger roughnesses compared to the inner layers.

Table IX(a): Parameters, such as layers, their thicknesses, ESLDs and surface/interface roughness obtained from the analysis of the XRR data.

| Layer | Thickness (Å) | ESLD ($10^{-5}$ Å$^{-2}$) | Roughness (Å) |
|---|---|---|---|
| CoO | 76±3 | 4.41±0.2 | 72±4 |
| Co | 155±6 | 6.27±0.14 | 27±3 |
| CoO | 82±5 | 4.41±0.13 | 17±2 |
| HD Co | 57±4 | 8.09±0.12 | 20±2 |
| Co | 191±6 | 6.32±0.15 | 3±1 |
| HD Co | 23±2 | 7.21±0.14 | 7±2 |
| Si sub | - | 2.08±0.05 | 8±1 |

Table IX(b): Parameters, such as layers, their thicknesses, ESLDs, MSLDs and surface/interface roughness obtained from the analysis of the PNR data.

| Layer | Thickness (Å) | NSLD ($10^{-6}$ Å$^{-2}$) | MSLD ($10^{-6}$ Å$^{-2}$) | Roughness (Å) |
|---|---|---|---|---|
| CoO | 75±3 | 3.93±0.14 | 0 | 41±7 |
| Co | 161±5 | 2.27±0.12 | 4.14±0.12 | 15±2 |
| CoO | 78±3 | 3.95±0.13 | 0 | 9±3 |
| HD Co | 54±4 | 2.91±0.12 | 0 | 10±2 |
| Co | 200±5 | 2.27±0.11 | 4.16±0.11 | 2.0±0.5 |
| HD Co | 24±2 | 2.6±0.10 | 0 | 5±1 |
| Si sub | - | 2.08±0.05 | 0 | 6±2 |

## 4. Conclusion

    In earlier experiments, where cobalt films were grown on various oxide substrates, no high density nonmagnetic cobalt was found. The absence of HD NM cobalt in our case, when the films are grown on oxides, are consistent with earlier results. For direct deposition of cobalt on silicon, epitaxial growth of cobalt grains at the Co/Si interface, and the consequent strain in these grains, is apparently responsible for the gorwth of HD NM cobalt layer at the Co/Si interface. For growth on oxides, the initial epitaxial grain growth would be absent. We also observe (in SIMS depth profile, not shown here) negligible cobalt silicide growth at the Co/Si interface. This is a possible indication of the reduced reactivity of the HD NM Co with Si.

For all thicknesses of the cobalt film, the same feature, namely HD NM layers at the Co/Si interface and near the top interface, namely CoO/Co or Au/Co, have been observed. HD NM Co layers are observed irrespective of whether the top surface of Co is Au-capped or it has a natural oxide. After the initial growth near the Co/Si interface, the growth is of Co on already deposited Co and at this stage during deposition diffusion into the grain boundaries is not dominant as the time for diffusion is apparently insufficient. When the dposition is stopped, the deposited atoms at the last stage has enough time for diffusing into the grain boundaries. This is the likely reason for the formation of compressed grains at the top of the film. In-situ XRR and PNR experiments during growth of the thin film, also at different deposition rates, can shine further light on the growth mechanism of the HD cobalt.

## Acknowledgement

We thank Prof. S. Banerjee for giving access to his electron beam deposition facility. We acknowledge the help provided by Avijit Das for SIMS measurement.